\documentclass[aps,prb,twocolumn,superscriptaddress,showpacs,floatfix]{revtex4}

\usepackage{graphicx}
\usepackage{bm}

\begin{document}

\title{Critical state in type-II superconductors of arbitrary shape}

\author{G.~P.~Mikitik}
\affiliation{Max-Planck-Institut f\"ur Metallforschung,
  D-70506 Stuttgart, Germany}
\affiliation{B.~Verkin Institute for Low Temperature Physics
  \& Engineering, National Ukrainian Academy of Sciences,
  Kharkov 61103, Ukraine}
\author{E.~H.~Brandt}
\affiliation{Max-Planck-Institut f\"ur Metallforschung,
  D-70506 Stuttgart, Germany}

\date{\today}

\begin{abstract}
The well-known Bean critical state equations in general are not
sufficient to describe the critical state of type-II
superconductors when the sample shape is not symmetric. We show
how one can find the critical state in superconductors of
arbitrary shape. Analyzing a simple example of nonsymmetry, we
demonstrate that in the general case, a perturbation of the
current distribution in the critical state propagates into the
sample smoothly in a diffusive way. This is in contrast to the
usual Bean critical state where the current distribution changes
abruptly at a narrow front.
\end{abstract}


\maketitle


The concept of the critical state introduced by Charles Bean
\cite{1} is widely used to describe various physical phenomena in
the vortex phase of type-II superconductors, see, e.g.,
Refs.~\onlinecite{2,3} and citations therein.
According to Bean, in the critical state of type-II
superconductors with flux-line pinning, the driving force of the
currents flowing in this state is balanced by the pinning force
acting on the vortices. The critical state is characterized by the
component of the current density flowing {\it perpendicular} to
the flux lines, $j_{c\perp}$, since only this component generates
a driving force. It is assumed in the critical-state theory that
this $j_{c\perp}$ is known, i.e., it is a given function of the
magnetic induction ${\bf B}$, $j_{c\perp}=j_{c\perp}({\bf B})$,
and the problem of this theory is to find the appropriate
distribution of the magnetic fields and currents in the critical
state. Below, to explain the physics with the least mathematical
complications, we shall imply the simplest form: $j_{c\perp}=$
const., which is frequently used in practice, but an extension to
the general case is straightforward. For simplicity, we also
assume that the magnetic fields ${\bf H}$ in the superconductor
considerably exceed the lower critical field $H_{c1}$, and so we
may put ${\bf B}=\mu_0 {\bf H}$.

If in the critical state a current-density component
$j_{\parallel}$ {\it parallel} to the local magnetic field is also
generated, the magnitude of $j_{\parallel}$ remains undefined, and
one thus cannot in general find the distributions of the magnetic
field ${\bf H}({\bf r})$ and current density ${\bf j}({\bf r})$ in
the critical state. Indeed, to solve the Maxwell equations for
${\bf H}$,
\begin{eqnarray}\label{bin} 
{\rm rot}{\bf H}={\bf j},\ \ \  {\rm div}{\bf H}=0,
\end{eqnarray}
it is necessary to know the magnitude and direction of the
currents ${\bf j}({\bf r})$ in the sample. \cite{4} However, one
has only the {\it two} conditions:
\begin{equation}\label{bin1} 
j_{\perp}=j_{c\perp},\ \ \  {\rm div}{\bf j}=0,
\end{equation}
for {\it three} quantities: $j_{\perp}$, $j_{\parallel}$ and the
angle defining the direction of $j_{\perp}$ in the plane normal to
${\bf H}({\bf r})$. Thus, the existing critical-state theory based
only on Eqs.~(\ref{bin}), (\ref{bin1})  is {\it not complete}.

The above Maxwell equations with conditions (\ref{bin1}) can
provide the description of the critical state when the shape of
the superconductor is sufficiently symmetric and the external
magnetic field is applied along a symmetry axis, so that some
constraint on the directions of the currents is known in advance.
For example, the direction of the currents is obvious for a slab
in an external magnetic field parallel to its surface. For an
infinitely long cylinder with arbitrary cross-section in a
magnetic field parallel to its axis, the currents flow
perpendicular to this axis, and the critical state problem is
solved. \cite{2} Another completely solvable case is infinitely
thin flat superconductors, \cite{3} for which the currents can
flow only in the plane. However, in the case, e.g., of a thin
rectangular platelet of {\it finite} thickness in a perpendicular
magnetic field, the above critical state equations are already
incomplete for determining the magnetic fields and currents in the
critical state. \cite{C1,6}

We emphasize that even for {\it simple} experimental situations
equations (\ref{bin}) together with conditions (\ref{bin1}) can be
insufficient for solving the critical-state problem. As an example
that we shall analyze below, consider an infinite slab of
thickness $d$. Let this slab fill the space $|x|$, $|y| <\infty $,
$|z|\le d/2$, and be in a constant and uniform external magnetic
field $H_a$ ($H_a\gg J_c\equiv j_{c\perp}d$) directed along the
$z$ axis, i.e., perpendicularly to the slab plane. Let then a
constant field $h_{ax}$ ($J_c/2\le h_{ax}\ll H_a$) be applied
along the $x$ axis, and after that the magnetic field $h_{ay}$
($h_{ay}\ll H_a$) is switched on in the $y$ direction. This
critical state problem is not fully defined. Indeed, the condition
${\rm div}{\bf j}=0$ yields $j_z=0$, i.e., the currents flow in
the $x$-$y$ planes. Then, to describe the critical state, we may
use the parametrization:
\begin{eqnarray*}
{\bf j}=j_{c}(\varphi,\theta,\psi) (\cos\varphi(z), \
\sin\varphi(z),0), \\
{\bf H}(z)={\bf H}_a+ {\bf h}(z), \\
{\bf h}(z)=(h_x(z), h_y(z),0),
\end{eqnarray*}
where $j_c(\varphi,\theta,\psi)$ is the magnitude of the critical
current density when a flux-line element is given by the angles
$\psi$ and $\theta$, $\tan\psi=h_y/h_x$, $\tan\theta =
(h_x^2+h_y^2)^{1/2}/H_a$, while the current flows in the direction
defined by the angle $\varphi$; all these angles generally depend
on $z$. A dependence of $j_c$ on the orientation of the local
${\bf H}$, $j_c(\varphi,\theta,\psi) = j_{c\perp}/[1 -
\cos^2(\varphi-\psi) \sin^2\theta]^{1/2}$,  \cite{6} appears if
${\bf j_c}$ is not perpendicular to this ${\bf H}$. However, at
$H_a\gg h_{ax}$, $h_{ay}$, $J_c$, the field ${\bf H}$ is
practically normal to the $x$-$y$ planes where the currents flow
($\theta \approx 0$), and we may put $j_c(\varphi,\theta,\psi) =
j_{c\perp}$. With this parametrization, the equation ${\rm
div}{\bf H}=0$ is satisfied identically, while the Maxwell
equation ${\rm rot}{\bf H}={\bf j}$ reads
\begin{eqnarray} \label{1}   
   {d h_x\over dz}=j_{c\perp}\sin\varphi , \\
   -{d h_y \over dz}=j_{c\perp}\cos\varphi , \label{2}
\end{eqnarray}
and one has only {\it two} equations for the {\it three} functions
$h_x(z)$, $h_y(z)$, $\varphi(z)$.

In real samples of nonsymmetric shape, adjacent flux lines may be
slightly rotated relative to each other in the critical state. It
is this rotation that generates a component of the current along
the magnetic field. The rotation of flux-lines can lead to their
mutual cutting. \cite{2,7} Flux line cutting occurs when the
component of the current density parallel to the magnetic field,
$j_{\parallel}$, exceeds some longitudinal critical current
density $j_{c\parallel}$. In this situation a vortex \cite{8} or a
vortex array \cite{9} becomes unstable with respect to a helical
distortion, and the growth of this distortion leads to flux-line
cutting. When both $j_{\parallel}$ and $j_{c\perp}$ are equal to
their critical values $j_{c\parallel}$ and $j_{c\perp}$,
respectively, Clem's double critical state \cite{7,Clem} occurs in
the superconductor. In this case one has three conditions for the
three quantities, and equations (\ref{bin}) are sufficient to
describe the double critical state. However, in many real
situations, $j_{\parallel}$ does {\it not} reach $j_{c\parallel}$,
and flux cutting then does {\it not} occur in the critical state.
It is such situations that we consider here. In particular, in the
above example the projection of the current density on the
magnetic field (which practically coincides with the $z$ axis) is
negligible, and flux cutting does not occur.


We now show how the critical state problem can be solved for
superconductors of arbitrary shape. Let the critical state be
known at some moment of time $t$, i.e., one has ${\bf H}=H({\bf
r}){\bm \nu}({\bf r})$ inside the superconductor where the
magnitude of the magnetic field, $H$, and the unit vector ${\bm
\nu}$ are both known functions of the coordinates ${\bf r}$ at
some external magnetic field ${\bf H}_a(t)$. The current density
${\bf j}({\bf r})$ in the critical state follows from the Maxwell
equation ${\bf j}={\rm rot}\left(H({\bf r}){\bm \nu}({\bf
r})\right)$, while the component of the current density
perpendicular to the magnetic field is given by ${\bf
j}_{\perp}={\bf j}- {\bm \nu}({\bm \nu}{\bf j})\equiv
j_{c\perp}{\bf n}_{\perp}({\bf r})$. Here the last equality
defines the unit vector ${\bf n}_{\perp}$. Let the external field
infinitesimally (and slowly \cite{C1a}) change by $\delta {\bf
H}_a= \dot {\bf H}_a \delta t$. We now shall find the new critical
state at the new external magnetic field ${\bf H}_a+\delta {\bf
H}_a$.

Under the change of ${\bf H}_a$, the critical currents locally
shift the vortices in the direction \cite{C2} of the Lorentz force
$[{\bf j}\times {\bm \nu}]$; this shift generates an electric
field directed along $[{\bm \nu}\times [{\bf j}\times {\bm
\nu}]]={\bf j}_{\perp}$, i.e., along the vector ${\bf n}_{\perp}$.
Thus, we can represent the electric field ${\bf E}({\bf r})$ in
the form ${\bf E}={\bf n}_{\perp}e$ where the scalar function
$e({\bf r})$ is the modulus of the electric field. Note that in
contrast to the Bean assumption, \cite{11} the electric field
generally is not parallel to the total current density ${\bf
j}({\bf r})$. Using the Maxwell equation
\begin{equation}\label{3}   
{\rm rot}(e {\bf n}_{\perp})=-\mu_0 \dot {\bf H},
\end{equation}
where $\dot {\bf H}\equiv \partial {\bf H}/ \partial t$, and the
equation
\begin{equation}\label{4}   
{\rm rot}\dot {\bf H}={\partial {\bf j}\over \partial t},
\end{equation}
one can express the change of the magnetic fields and currents via
{\it one scalar} function $e({\bf r})$. This function can be found
from the condition that in the critical state the absolute value
of ${\bf j}_{\perp}$ is a given function of ${\bf B}$, $j_{\perp}=
j_{c\perp}({\bf B})$, or in the differential form,
\[
 {\bf j}_{\perp}\!\cdot {\partial {\bf j}_{\perp}\over \partial t}
 =j_{c\perp}({\bf B})\left ( {\partial j_{c\perp}({\bf B}) \over
 \partial {\bf B}} \cdot \mu_0 \dot {\bf H}\right ).
\]
In our case when $j_{c\perp}=$const., this condition reads
\[
 {\bf j}_{\perp}\!\cdot {\partial {\bf j}_{\perp}\over \partial t}
 =0.
\]
Taking into account the definition of ${\bf j}_{\perp}$ and using
the identities
\begin{eqnarray*}
 H\dot {\bm \nu}= \dot {\bf H}- ({\bm \nu}\cdot \dot {\bf
 H}){\bm \nu}, \\
 H \dot {\bm \nu}\cdot {\bf j}=\dot {\bf H}\cdot
 {\bf j}_{\perp}, \\
 {\bf j}\cdot {\bm \nu}= ({\bm \nu}\cdot {\rm rot}{\bm \nu})H,
\end{eqnarray*}
we arrive at an equation for $e({\bf r})$,
\begin{equation}\label{5}   
{\bf n}_{\perp}\!\cdot \{ {\rm rot\, rot}(e{\bf n}_{\perp})- ({\bm
\nu}\cdot {\rm rot}{\bm \nu})\, {\rm rot}(e{\bf n}_{\perp}) \}=0.
\end{equation}
Continuity of the magnetic field on the surface of the
superconductor, $S$, yields the boundary condition:
  \begin{eqnarray}\label{6}   
  -{\rm rot}\left (e({\bf r}_S){\bf n}_{\perp}({\bf r}_S )\right )
  = \mu_0 \dot {\bf H}_a + ~~~~~~~~~~~~  \nonumber \\
  ~~~~~~~  \int {[{\bf R}\times {\rm rot\, rot}(e({\bf
  r'}){\bf n}_{\perp}({\bf r'}))] \over 4\pi R^3}\, d {\bf r}',
  \end{eqnarray}
where ${\bf r}_S$ is a point on the surface $S$, ${\bf R}\equiv
{\bf r}_S-{\bf r}'$, $R=|{\bf R}|$, and the integration is carried
out over the volume of the sample. The right hand side of this
boundary condition expresses $\mu_0 \dot{\bf H}$ on the surface of
the superconductor (but reaching from outside) with the use of the
Biot-Savart law. If in the critical state of the superconductor
there are also boundaries at which the direction of the critical
currents changes discontinuously or which separate regions with
$j_{\perp}=j_{c\perp}$ from regions with $j=0$, \cite{C3} the
function $e({\bf r})$ has to vanish at these boundaries.
Otherwise, the electric field $e{\bf n}_{\perp}$ would be
discontinuous there.

After determining the function $e({\bf r})$, one can find the new
critical state ${\bf H}({\bf r})+\delta {\bf H}({\bf r})$ using
the definition $\delta {\bf H}({\bf r})= \dot {\bf H}\delta t$ and
Eq.~(\ref{3}). We emphasized that the new critical state depends
only on the previous state ${\bf H}({\bf r})$ and on the change of
the external field $\delta {\bf H}_a= \dot {\bf H}_a \delta t$.
The dependence on $\delta {\bf H}_a$ follows from the
proportionality of $e$, $\dot {\bf H}$, $\partial {\bf j}_{\perp}/
\partial t$ to $\dot {\bf H}_a$, which results from the linearity
of Eqs.~(\ref{3})-(\ref{6}). Note that in agreement with the
meaning of the critical state, the new state will be the same for
different sweep rates of the external magnetic field, $\dot {\bf
H}_a$, since it depends only on the {\it product} $\dot {\bf H}_a
\delta t=\delta {\bf H}_a$. On the other hand, the electric field
$e$ plays an auxiliary role in the above description since it is
proportional to $\dot {\bf H}_a$ rather than to $\delta {\bf
H}_a$. The presented description also shows that the critical
state generally depends on the history of its creation. In other
words, it depends not only on the final value of the external
magnetic field ${\bf H}_a$ but also on the sequence of steps
$\delta {\bf H}_a$ that lead to this value.

The above approach, which is in essence the generalization of the
appropriate analyses used for a slab, \cite{Clem,G1} can be summed
up as follows: We add to the static equations (\ref{bin}) the
quasistatic Maxwell equation (\ref{3}). It is known \cite{4} that
for this set of the equations to be solvable, it has to be
supplemented by some law ${\bf E}({\bf j})$. We introduce this law
from well-known physical ideas: At any given ${\bf j}$ and ${\bf
B}$ (determined by the previous critical state), the {\it
direction} of ${\bf E}$ follows from the formula ${\bf E}=[{\bf
B}\times {\bf v}]$ where ${\bf v}$ is the vortex velocity caused
by the Lorentz force $[{\bf j}\times {\bf B}]$. As to the {\it
magnitude} of ${\bf E}$, it is found from the condition that
$|{\bf j}_{\perp}|=j_{c\perp}$. In fact, this condition may be
interpreted as a current--voltage law with $|{\bf E}|=e=0$ at
$j_{\perp}<j_{c\perp}$ and $e\to \infty$ at
$j_{\perp}>j_{c\perp}$, which is usually implied in the
description of the ideal critical state. \cite{C1a}

Recently, \cite{BL1,BL2} a variational principle was put forward
to describe the critical state in superconductors. In deriving
this principle Bad\'ia and L\'opez used Eqs.~(\ref{bin}),
(\ref{3}) and a current--voltage law with $|{\bf E}|=0$ at
$j<j_{c}$ and $|{\bf E}|\to \infty$ at $j>j_c$. However, the
physical idea on the {\it direction} of the electric field was not
incorporated in their theory. For some situations this leads to
contradiction with existing concepts.\cite{8,9}  In particular,
in their so-called isotropic model with $H$-independent $j_c$, the
electric field ${\bf E}$ is parallel to ${\bf j}$, and hence a
nonzero ${\bf E}$ along ${\bf H}$ appears even for an
infinitesimal longitudinal component of ${\bf j}$, i.e.,
flux-cutting occurs without any threshold $j_{c\parallel}$.


 \begin{figure}  
\includegraphics[scale=.5]{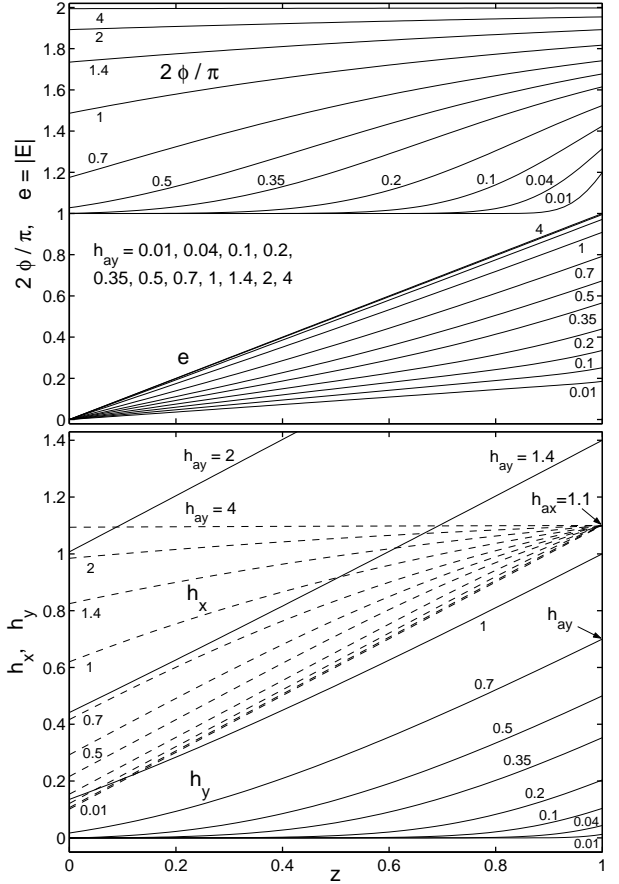}
\caption{\label{fig1}Profiles of the angle of currents
$\varphi(z)$, magnitude of electric field $e(z)$ (top) and
magnetic field components $h_x(z)$ (dashed lines), $h_y(z)$ (solid
lines) (bottom) in the critical states of the slab described by
Eqs.~(\ref{1}), (\ref{2}), (\ref{e}), (\ref{phi}), (\ref{b1}) -
(\ref{in});  $h_{ax}=1.1$ and $h_{ay} = 0.01$, $0.04$, $0.1$,
$0.2$, $0.35$, $0.5$, $0.7$, $1$, $1.4$, $2$, $4$. We start at
$h_{ay} = 0$ with $h_x(z)=h_{ax}-1+z$, $h_y(z)=0$, and $\phi(z) =
\pi/2$. Here $z$ is in units $d/2$, $h_x$ and $h_y$ in units
$j_{c\perp}d/2 = J_c/2$, and $e$ in units $\mu_0(d h_{ay}/d t)d/2$.
 } \end{figure}  

To illustrate the obtained results, we now consider the example
mentioned above. In the case of the slab, equation (\ref{5}) for
the electric field $e$ takes the form:
\begin{equation}\label{e}   
 e''-(\varphi')^2 e=0 ,
\end{equation}
where $\varphi' \equiv \partial \varphi/\partial z$ and $e''\equiv
\partial^2 e /\partial z^2$. For the angle $\varphi$ we obtain
from Eqs.~(\ref{3}) and (\ref{4}):
\begin{equation}\label{phi} 
 \mu_0 j_{c\perp}{\partial \varphi \over \partial t}=2e'\varphi'+
 e\varphi'' .
\end{equation}
These equations complement Eqs.~(\ref{1}), ({\ref{2}), and now we
have {\it four equations for four functions}. The boundary
conditions to Eqs.~(\ref{1}), (\ref{2}), (\ref{e}), (\ref{phi})
at $z=d/2$ are
\begin{eqnarray}\label{b1}  
  h_x=h_{ax},\ \ \  h_y=h_{ay}(t) , \\
  (e\cos\varphi)'=-\mu_0 {dh_{ay}(t)\over dt},\ \ \
  (e\sin\varphi)'=0,  \label{b2}
\end{eqnarray}
or equivalently, conditions (\ref{b2}), which follow from formula
(\ref{6}), can be rewritten in the form:
\begin{equation}\label{b2a} 
  e'=-\mu_0 {dh_{ay}(t)\over dt}\cos\varphi,\ \ \
  e\varphi'=\mu_0 {dh_{ay}(t)\over dt}\sin\varphi.
\end{equation}
Taking into account the symmetry of the problem, \cite{C4} it is
sufficient to solve equations (\ref{1}), (\ref{2}), (\ref{e}),
(\ref{phi}) in the region $0 \le z \le d/2$. At $z=0$, where the
direction of the currents changes discontinuously, one has the
additional condition for $e$,
\begin{equation}\label{b4}  
  e(0)=0.
\end{equation}
Since after switching on $h_{ax}$, the critical currents flow in
the $y$ direction, we have the following initial state for
Eq.~(\ref{phi}):
\begin{equation}\label{in}  
  \varphi (z,t=0)=\pi/2 ,
\end{equation}
where the moment $t=0$ corresponds to the beginning of switching
on $h_{ay}$.

In agreement with the general considerations given above, it
follows from Eq.~(\ref{phi}) and $e\propto d h_{ay}/d t$ that the
angle  $\varphi$ is a function of $z$ and $h_{ay}$ rather than of
$z$ and $t$. In fact, this equation describes $\varphi(z)$ in the
sequence of the critical states developed in the process of
increasing $h_{ay}$. The solution of equations (\ref{1}),
(\ref{2}), (\ref{e}), (\ref{phi}) with conditions (\ref{b1}) -
(\ref{in}) is shown in Fig.~1. Interestingly, when $h_{ay}$
increases, the {\it other component} of the magnetic field, $h_x$,
penetrates further into the slab [at $h_{ay}\sim J_c\equiv
j_{c\perp}d$, $h_x(z)$ almost coincides with $h_{ax}$], and the
angle $\varphi$ tends to $\pi$. In other words, with increasing
$h_{ay}$ the initial critical state for the component $h_x$
relaxes, while the critical state for $h_y$ is developed. Note
that we should arrive at a different critical state with $\varphi=
\pi/2+ {\rm arctan}(h_{ay}/h_{ax})$ if the $x$ and $y$ components
of the external field were increased simultaneously
[$h_{ay}(t)/h_{ax}(t)=$const.] from zero to the same values
$h_{ax}$, $h_{ay}\sim J_c$. Thus, the dependence of the critical
state on its prehistory is clearly seen even in this simple
example.

Figure 1 also reveals the following two interesting features of
the critical state: (a) The visible penetrating front of $h_y$
reaches the center of the slab when $h_{ay}$ is still less than
$J_c/2$, the field of full penetration in the Bean case. (b) The
change of the angle $\varphi(z,h_{ay})$ has {\it diffusive
character\/}. This is in stark contrast to the usual Bean critical
state, in which any change of the current direction occurs inside
a narrow front.

Interestingly, equations (\ref{1}), (\ref{2}), (\ref{e}),
(\ref{phi}) are applicable also to a number of other physical
problems if the boundary and initial conditions are changed
appropriately. In particular, these equations also describe the
usual Bean critical state in the slab, corresponding to a {\it
discontinuous} solution $\varphi(z)$. Using these equations, one
can also investigate the low-frequency response of the slab to a
{\it circularly} polarized ac field applied in the plane of the
sample perpendicularly to the large magnetic field $H_a$.
\cite{G2} It is clear from the data of Fig.~1 that this response
will differ from the response to a linearly polarized ac field,
for which the analysis based on the usual Bean model is
applicable. These equations also enable one to consider the
vortex-shaking effect: \cite{14} If the field $H_a$ is not uniform
in the plane of the slab, and thus a sheet current ${\bf J}$ flows
in it, a small ac field applied along the current leads to a
continuous drift of vortices in the direction $[{\bf J}\times {\bf
H}_a]$. It turns out that Eqs.~(\ref{1}), (\ref{2}), (\ref{e}),
(\ref{phi}) have a solution that reproduces this result of
Ref.~\onlinecite{14}, obtained there by a different method using
geometrical arguments.

In summary, we have extended the critical-state theory to the
general case when the sample is not sufficiently symmetric, or
when the external field is not along a symmetry axis or its
direction changes in some complex manner. In such situations the
currents in the critical state need not be perpendicular to the
local magnetic fields, and a {\it longitudinal component} of the
currents with respect to these fields exists in the
superconductor. When the magnitude of the longitudinal current
density $j_{\parallel}$ does not exceed some critical value
$j_{c\parallel}({\bf B})$, the critical state can be found using
our Eqs.~(\ref{3}) - (\ref{6}). Such a state, in general,
essentially differs from the usual Bean critical state. In other
words, the Bean state is only a {\it special case} of the general
critical state. When the component $j_{\parallel}$ reaches
$j_{c\parallel}$ in the sample (or in part of its volume), Clem's
double critical state develops there. \cite{7, Clem}


  This work was supported by the German Israeli Research Grant
Agreement (GIF) No G-705-50.14/01 and by the INTAS project No
01-2282.


\begin{thebibliography}{1}
\bibitem{C1} This is due to the fact that although currents are still
parallel to the plane of the platelet, in principle they can have
different directions at different depths. The critical-state
problem in the special case of thin flat superconductors in a
perpendicular magnetic field was solved in Ref.~\onlinecite{6}.
\bibitem{C2} In the case of anisotropic pinning the direction
of the shift may differ from $[{\bf j}\times {\bm \nu}]$, see
Ref.~\onlinecite{6}. But it is important here that even in this
case, the direction of the shift can be expressed \cite{6} via the
directions of ${\bf j}$ and ${\bm \nu}$. Below, to avoid
mathematical complications, we consider the simplest case.
\bibitem{C3} When ${\bf H}_a$ changes, these boundaries may shift
in the superconductor. New positions of the boundaries are found
from continuity of ${\bf H}$ in the sample.
\end{thebibliography}
\end{document}